\documentclass[11pt,letterpaper]{article}
\usepackage{emnlp2017}
\usepackage{times}
\usepackage{latexsym}
\usepackage{amsmath,amsfonts,amssymb}
\usepackage{eqnarray}
\usepackage{mathtools}
\usepackage{balance} 
\usepackage{mathrsfs}
\usepackage{multirow}
\usepackage{csquotes}
\usepackage{bbm}
\usepackage{enumitem} 

\usepackage[]{algorithm2e}
\usepackage{graphics}
\usepackage{subfigure}
\usepackage{units}
\usepackage{booktabs}

\emnlpfinalcopy

\def\emnlppaperid{***}

\newcommand{\paragraphHdNospace}[1] {\noindent\textbf{#1.}} 
\newcommand{\paragraphHd}[1] {\vspace{1.2mm}\noindent\textbf{#1.}} 
\newcommand{\paragraphSc}[1] {\vspace{1.2mm}\noindent\textsc{#1.}} 
\newcommand{\ra}{\renewcommand{\arraystretch}{1.2}}

\makeatletter
\def\@fnsymbol#1{\ensuremath{\ifcase#1\or *\or \dagger\or \ddagger\or
   \mathsection\or \mathparagraph\or \|\or **\or \dagger\dagger
   \or \ddagger\ddagger \else\@ctrerr\fi}}
\makeatother
 
\newcommand{\specialcell}[2][c]{%
  \begin{tabular}[#1]{@{}c@{}}#2\end{tabular}}

\title{PACRR: A Position-Aware Neural IR Model for Relevance Matching}

\author{Kai Hui\\MPI for Informatics\\ Saarbr\"ucken \\Graduate School of \\Computer Science
\And
Andrew Yates\\MPI for Informatics
\And
Klaus Berberich\\ htw saar \\MPI for Informatics
\And
Gerard de Melo\\Rutgers University\\New Brunswick} 

\begin{document}
\maketitle

\begin{abstract}
In order to adopt deep learning for information retrieval, models are needed that can capture all relevant information required to assess the relevance of a document to a given
user query. While previous
works have successfully captured unigram term matches, how to fully employ 
position-dependent information such as proximity and term dependencies
has been insufficiently explored. 
In this work,
we propose a novel neural IR model named \textit{PACRR}
aiming at better modeling position-dependent interactions between a 
query and a document.
Extensive experiments on six years' \textsc{Trec} Web Track data confirm that the proposed
model yields better results under multiple benchmarks.
\end{abstract}

\section{Introduction} 
\label{sec.intro}

Despite the widespread use of deep neural models across a range of linguistic tasks,
to what extent such models can improve information retrieval (IR) and
which components a deep neural model for IR should include remain open questions.
In ad-hoc IR, 
the goal is to produce a ranking of relevant documents given an open-domain (``ad hoc'') query and a document collection.
A ranking model thus aims at evaluating the interactions between
different documents and a query, assigning higher scores to documents that
better match the query.
Learning to rank models, like the recent IRGAN model~\cite{wang2017irgan},
rely on handcrafted features to encode query document interactions,
e.g., the relevance scores from unsupervised ranking models.
Neural IR models differ in that they extract interactions 
directly based on the queries and documents.
Many early neural IR models can be categorized as \emph{semantic matching} models, 
as they embed both queries and documents into
a low-dimensional space, and then assess their similarity based on such dense
representations. Examples in this regard include 
\textit{DSSM}~\cite{Huang:2013:LDS:2505515.2505665}
and \textit{DESM}~\cite{mitra2016dual}.
The notion of relevance is inherently asymmetric, however, 
making it different from well-studied semantic matching tasks such as semantic relatedness and paraphrase detection. 
Instead, \emph{relevance matching} models such as 
MatchPyramid~\cite{DBLP:journals/corr/PangLGXC16},
DRMM~\cite{guo2016deep} 
and the recent K-NRM~\cite{xiong2017end}
resemble traditional IR retrieval measures in that they directly consider
the relevance of documents' contents with respect to the query.
The DUET model~\cite{mitra2017learning} 
is a hybrid approach that combines signals from a local model for relevance matching
and a distributed model for semantic matching.  
The two classes of models are fairly distinct. In this work, we focus 
on relevance matching models.

Given that relevance matching approaches mirror ideas from
traditional retrieval models, 
the decades of research on ad-hoc IR
can guide us with regard to the specific kinds of 
relevance signals a model ought to capture.
Unigram matches are the most obvious signals to be 
modeled, as a counterpart to the term frequencies that appear in 
almost all traditional retrieval models.
Beyond this,
positional information, including
where query terms occur
and how they depend on each other, can also be exploited, as demonstrated in
retrieval models that are aware of term proximity~\cite{tao2007exploration} and
term dependencies~\cite{DBLP:conf/cikm/HustonC14,metzler2005markov}.
Query coverage is another factor that can be used to ensure that,
for queries with multiple terms,
top-ranked documents contain multiple query terms rather than emphasizing only one query term.
For example, given the query ``dog adoption requirements'', unigram matching signals
correspond to the occurrences of the individual terms ``dog'', ``adoption'', or
``requirements''. When considering positional information,
text passages with ``dog adoption'' or ``requirements for dog adoption'' 
are highlighted, distinguishing them from 
text that only includes individual terms.
Query coverage, meanwhile, further emphasizes that matching signals 
for ``dog'', ``adoption'', and  ``requirements'' should all be included in a document.

Similarity signals from unigram matches are taken as input 
by DRMM~\cite{guo2016deep} after being summarized as histograms,
whereas K-NRM~\cite{xiong2017end} directly digests
a query-document similarity matrix and summarizes it with multiple kernel functions.
As for positional information,
both the MatchPyramid~\cite{DBLP:journals/corr/PangLGXC16} and local DUET~\cite{mitra2017learning} models account for it by
incorporating convolutional layers based on similarity matrices between queries and documents.
Although this leads to more complex models, both have difficulty in significantly
outperforming the DRMM model~\cite{guo2016deep,mitra2017learning}.
This indicates that it is non-trivial to
go beyond unigrams by utilizing positional
information in deep neural IR models.
Intuitively,
unlike in standard sequence-based models, the interactions between a query and a document are sequential along the query axis as well as along the document axis, 
making the problem multi-dimensional in nature. 
In addition, this makes it non-trivial to combine matching signals from
different parts of the documents and over different query terms.
In fact, we argue that both MatchPyramid and local DUET models 
fail to fully account for one or more of the aforementioned factors.
For example, as a pioneering work, MatchPyramid
is mainly motivated by models developed in 
computer vision, resulting in 
its disregard of certain IR-specific considerations
in the design of components, 
such as pooling sizes that ignore the query and document dimensions.
Meanwhile, local DUET's CNN filters match entire documents against individual query terms, neglecting proximity and possible dependencies among different query terms.

We conjecture that a suitable combination of convolutional kernels and recurrent layers can lead to a model that better accounts for these factors.
In particular,
we present a novel re-ranking model called \textit{PACRR}
(\emph{Position-Aware Convolutional-Recurrent Relevance} Matching).
Our approach first produces similarity matrices that record the semantic
similarity between each query term and each individual term occurring
in a document. These matrices are then fed through a series of convolutional, max-k-pooling, and recurrent layers so as to capture interactions corresponding to, for instance, bigram and trigram matches, and finally to aggregate the signals in order to produce global relevance assessments.
In our model, the convolutional layers are designed to capture both unigram matching and 
positional information over text windows with different lengths;  
k-max pooling layers are along the query dimension, preserving matching signals 
over different query terms; the recurrent layer combines
signals from different query terms to produce a query-document relevance score.

\paragraphHdNospace{Organization} The rest of this paper unfolds as
follows. 
Section~\ref{sec.method} describes our approach for computing
similarity matrices and the architecture of our deep learning
model. The setup and results of our extensive experimental evaluation
can be found in Section~\ref{sec.eval}, before
concluding in Section~\ref{sec:conclusion}.


\section{The PACRR Model}\label{sec.method}

\begin{figure*}
 \centering   
 \includegraphics[width=\linewidth]{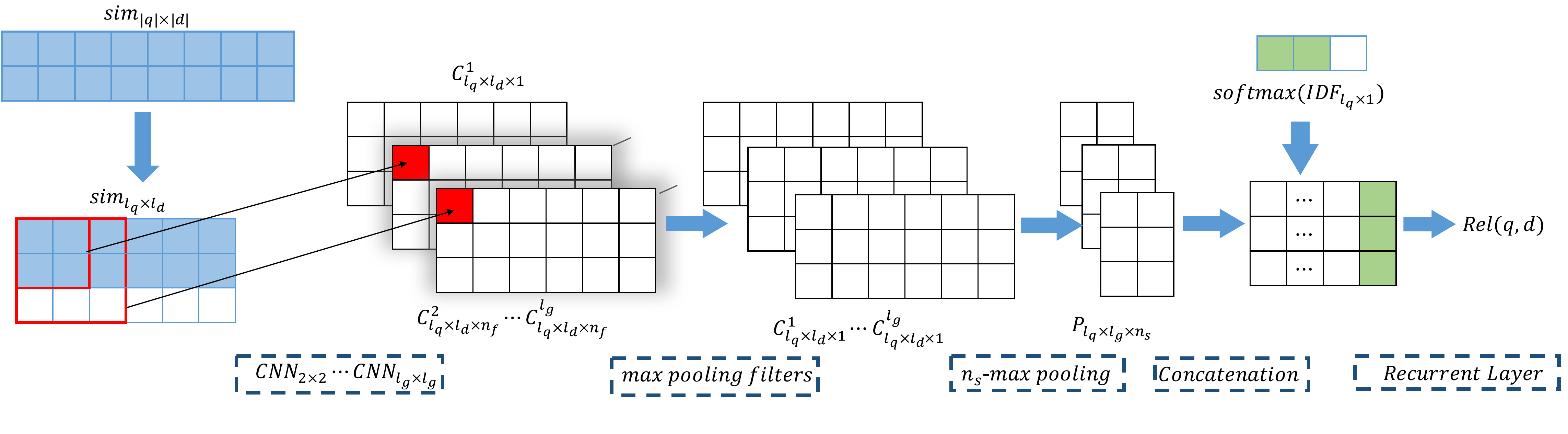}
 \caption{The pipeline of \textit{PACRR}. Each
   query $q$ and document $d$ is first converted
   into a query-document similarity matrix $\mathit{sim}_{|q|\times |d|}$.
   Thereafter, a distillation method (\textit{firstk} is displayed) transforms the raw similarity matrix into unified dimensions, namely,
   $\mathit{sim}_{l_q\times l_d}$. 
   Here, $l_g-1$ convolutional layers (CNN) are applied to the distilled similarity matrices. 
   As $l_g=3$ is shown, layers with kernel size $2$ and $3$ are applied.
   Next, max pooling is applied, leading to $l_g$ matrices $C^1\cdots C^{l_g}$.
   Following this,
   $n_s$-max pooling captures the strongest $n_s$ signals over each query term and n-gram size, and
   the case for $n_s=2$ is shown here.
   Finally, the similarity signals from different n-gram sizes are concatenated, the query terms'
   normalized IDFs are added, and a recurrent layer combines
   these signals for each query term into a query-document relevance score $\mathit{rel}(q,d)$.
 }
    \label{fig.model}
 \end{figure*}

We now describe our proposed \textit{PACRR} approach, which 
consists of two main parts: a relevance matching
component that converts each query-document pair into a similarity matrix $\mathit{sim}_{|q|\times |d|}$ and
a deep architecture that takes a given query-document similarity matrix as input and produces a
query-document relevance score $\mathit{rel}(q,d)$.
Note that in principle the proposed model can be trained end-to-end by backpropagating
through the word embeddings, as in~\cite{xiong2017end}.
In this work, however,
we focus on highlighting the building blocks aiming at capturing positional information,
and freeze the word embedding layer to
achieve better efficiency.
The pipeline is summarized in Figure~\ref{fig.model}.

\subsection{Relevance Matching}
We first encode the query-document relevance matching via query-document
similarity matrices $\mathit{sim}_{|q|\times |d|}$ that encodes the similarity
between terms from a query $q$ and a document $d$,
where $\mathit{sim}_{ij}$  
corresponds to the similarity between the $i$-th term 
from $q$ and the $j$-th term from $d$.
When using cosine similarity, we have $\mathit{sim}\in [-1,1]^{|q| \times |d|}$.
As suggested in~\cite{hui2017position}, query-document similarity matrices
preserve a rich signal that can be used to perform relevance matching beyond unigram matches.
In particular, n-gram matching corresponds to consecutive 
document terms that are highly similar to at least one of the
query terms.
Query coverage is reflected in the number of rows in $\mathit{sim}$ that include
at least one cell with high similarity. 
The similarity between a query term $q$ and document term $d$
is calculated by taking the cosine similarity 
using the pre-trained\footnote{\url{https://code.google.com/archive/p/word2vec/}}
\textit{word2vec}~\cite{mikolov2013distributed}.

The subsequent processing in PACRR's convolutional layers requires that each query-document
similarity matrix have the same dimensionality. Given that the lengths of queries and documents vary, we first transform the raw similarity matrices $\mathit{sim}_{|q|\times |d|}$ into $\mathit{sim}_{l_q\times l_d}$ matrices with uniform
$l_q$ and $l_d$ as the number of rows and columns.
We unify the query dimension $l_q$ by zero padding
it to the maximum query length. With regard to the document dimension $l_d$,
we describe two strategies: \textit{firstk} and \textit{kwindow}.

\paragraphHd{PACRR-firstk}
Akin to~\cite{mitra2017learning}, the \textit{firstk} distillation method
simply keeps the first $k$ columns in the matrix, which correspond to the first $k$ terms in the document.
If $k>|d|$, the remaining columns are zero padded.

\paragraphHd{PACRR-kwindow}
As suggested in~\cite{guo2016deep}, relevance matching is local.
Document terms that have a low query similarity relative to a document's other terms
cannot contribute substantially to the document's relevance score.
Thus relevance matching can be extracted in terms 
of pieces of text that include relevant information.
That is, one can segment documents according to relevance 
relative to the given query and retain only the
text that is highly relevant to the given query.
Given this observation, we prune query-document similarity cells with a low similarity score.
In the case of unigrams,  we simply choose the top $l_d$ terms with the highest similarity
to query terms.
In the case for \textit{text snippets} beyond length $n$, 
we produce a similarity matrix $\mathit{sim}^n_{l_q\times l_d}$ for each
query-document pair and each $n$, because $n$ consecutive terms must be co-considered later on.
For each text snippet with length $n$ in the document, 
\textit{kwindow} calculates the maximum similarity between each term and the query terms, and then
calculates the average similarity over each $n$-term window.
It then selects the top $k = \left \lfloor{l_d/n}\right \rfloor$ windows by 
averaging similarity and discards all other
terms in the document. The document dimension is zero padded if $\left \lfloor{l_d/n}\right \rfloor$ is not a multiple of $k$.
When the convolutional layer later operates on a similarity matrix produced by \textit{kwindow},
the model's stride is set to $n$ 
since it can consider at most $n$ consecutive terms that are present in the original document.
This variant's output is a similarity matrix $\mathit{sim}^n_{l_q\times l_d}$ for each size $n$.

\subsection{Deep Retrieval Model}
Given a query-document similarity matrix $\mathit{sim}_{l_q\times l_d}$ as input, our deep architecture relies on convolutional layers 
to match every text snippet with length $n$
in a query and in a document to produce
similarity signals for different $n$.
Subsequently, two consecutive max pooling layers extract the document's strongest similarity cues for each $n$.
Finally, a recurrent layer aggregates these salient signals to predict a global query-document relevance score $\mathit{rel}(q,d)$.

\paragraphHd{Convolutional relevance matching over local text snippets}
The purpose of this step is to match 
text snippets with different length
from a query and a document given
their query-document similarity matrix as input.
This is accomplished by applying
multiple two-dimensional convolutional layers with different kernel sizes
to the input similarity matrix.
Each convolutional layer is responsible for a specific $n$; by
applying its kernel on $n\times n$ windows, it produces
a similarity signal for each window.
When the \textit{firstk} method is used, each convolutional layer receives the same similarity matrix
$\mathit{sim}_{l_q\times l_d}$ as input because \textit{firstk} produces the same similarity matrix
regardless of the $n$. When the \textit{kwindow} method is used, each convolutional layer
receives a similarity matrix $\mathit{sim}^n_{l_q\times l_d}$ corresponding to 
the convolutional layer with a $n\times n$ kernel.
We use $l_{g}-1$ different convolutional layers 
with kernel sizes $2\times2, 3\times 3, \dots, l_{g}\times l_{g}$, 
corresponding to bi-gram, tri-gram, \dots, $l_{g}$-gram matching, respectively, where the
length of the longest text snippet to consider is governed by a hyper-parameter
$l_{g}$.
The original similarity matrix corresponds to unigram matching, while
a convolutional layer with kernel size $n\times n$ is responsible for capturing
matching signals on $n$-term text snippets.
Each convolutional layer applies $n_f$ different filters to its input, where
$n_f$ is another hyper-parameter.
We use a stride of size (1, 1) for the \textit{firstk} distillation method,
meaning that the convolutional kernel advances one step at a time in both the query and document dimensions.
For the \textit{kwindow} distillation method, we use a stride of (1, $n$) to move the convolutional kernel one
step at a time in the query dimension, but $n$ steps at a time in the document dimension.
This ensures that the convolutional kernel only operates over consecutive terms that existed in the original document.
Thus, we end up with $l_{g}-1$ matrices $\mathcal{C}^n_{l_q\times l_d\times n_f}$,
and
the original similarity matrix is directly employed to handle the signals over unigrams.

\paragraphHd{Two max pooling layers}
The purpose of this step is to capture the $n_s$ strongest similarity signals for each query term.
Measuring the similarity signals separately for each query term allows the model to consider query term coverage,
while capturing the $n_s$ strongest similarity signals for each query term 
allows the model to consider signals from different kinds of relevance matching patterns, e.g., 
n-gram matching and non-contiguous matching. 
In practice, we use a small $n_s$ to prevent the model from being
biased by document length;
while each similarity matrix contains the same number of document term scores,
longer documents have more opportunity to contain terms that are similar to query terms.
To capture the strongest $n_s$ similarity signals for each query term,
we first perform max pooling over the filter dimension $n_f$ to keep only the strongest signal from the $n_f$ different filters,
assuming that there only exists one particular true matching pattern in a 
given $n\times n$ window, 
which serves different purposes compared with other tasks, such as the sub-sampling in
computer vision.
We then perform k-max pooling \cite{kalchbrenner2014convolutional} over the query dimension $l_q$
to keep the strongest $n_s$ similarity signals for each query term.
Both pooling steps are performed
on each of the $l_{g}-1$ matrices $\mathcal{C}^i$ from the convolutional layer
and on the original similarity matrix, which captures unigram matching, to
produce the 3-dimensional tensor $\mathcal{P}_{l_q\times l_g\times n_s}$.
This tensor contains the $n_s$ strongest signals for each query term and for each n-gram size across all $n_f$ filters.

\paragraphHd{Recurrent layer for global relevance}
Finally, our model transforms the query term similarity signals in $\mathcal{P}_{l_q\times l_g\times n_s}$
into a single document relevance score $\mathit{rel}(q,d)$.
It achieves this by applying a recurrent layer to $\mathcal{P}$, taking a sequence of vectors
as input and learning weights to transform them into the final relevance score.
More precisely, akin to~\cite{guo2016deep},
the IDF of each query term $q_i$ is passed through a softmax layer for normalization.
Thereafter, we split up the query term dimension to produce a matrix $\mathcal{P}_{l_g\times n_s}$
for each query term $q_i$, subsequently forming 
the recurrent layer's input by flattening each matrix
$\mathcal{P}_{l_g\times n_s}$ into a vector by concatenating the matrix's rows together and appending query term $q_i$'s
normalized IDF onto the end of the vector. This sequence of vectors for each query term $q_i$ is passed
into a Long Short-Term Memory (LSTM) recurrent layer~\cite{Hochreiter1997} with an output dimensionality of one.
That is, the LSTM's input is a sequence of query term vectors where each vector is composed
of the query term's normalized IDF and the aforementioned salient signals for the query term along different kernel sizes.
The LSTM's output is then used as our document relevance score $\mathit{rel}(q,d)$.

\paragraphHd{Training objective}
Our model is trained on triples consisting of a query $q$, relevant document $d^+$, and non-relevant document $d^-$,
minimizing a standard pairwise max margin loss as in Eq.~\ref{eq.loss}.
\begin{equation}\label{eq.loss}
\scriptstyle
 \mathcal{L}(q,d^+, d^-;\Theta)=\mathit{max}(0,1-\mathit{rel}(q,d^+)+\mathit{rel}(q,d^-))
\end{equation}


\section{Evaluation} 
\label{sec.eval}

In this section, we empirically evaluate \textit{PACRR} models using manual 
relevance judgments from the standard \textsc{Trec} Web Track. 
We compare them against several state-of-the-art neural IR models\footnote{We also attempted to 
include IRGAN~\cite{wang2017irgan} model as a baseline, but failed to obtain reasonable results 
when training on \textsc{Trec} data.}, including \textit{DRMM}~\cite{guo2016deep},
DUET~\cite{mitra2017learning}, 
\textit{MatchPyramid}~\cite{DBLP:journals/corr/PangLGXC16},
and \textit{K-NRM}~\cite{xiong2017end}.
The comparisons are over three task settings:
re-ranking search results from a simple initial ranker (\textsc{RerankSimple});
re-ranking all runs from the \textsc{Trec} Web Track (\textsc{RerankALL});
and examining neural IR models' classification accuracy between
document pairs (\textsc{PairAccuracy}).

\subsection{Experimental Setup}\label{sec.expsetting}
We rely on the widely-used 2009--2014 \textsc{Trec} Web
Track ad-hoc task benchmarks\footnote{\url{http://trec.nist.gov/tracks.html}}.
The benchmarks are based on the \textsc{ClueWeb09} 
and \textsc{ClueWeb12} 
datasets as document collections.
In total, there are 300 queries and more than 100k judgments (qrels).
Three years (2012--14) of query-likelihood baselines\footnote{Terrier~\cite{ounis06terrier} 
version without filtering spam documents} provided by  
\textsc{Trec}\footnote{\url{https://github.com/trec-web/trec-web-2014}} 
serve as baseline runs in the \textsc{RerankSimple} benchmark. 
In the \textsc{RerankALL} setting,
the search results from 
runs submitted by participants from each year are also considered: there are
71 (2009), 55 (2010),
62 (2011), 48 (2012), 50 (2013), and 27 (2014) runs.
ERR@20~\cite{Chapelle2009ERR} and nDCG@20~\cite{jarvelin2002cumulatedNDCG}
are employed as evaluation measures, and both are computed with the script from 
\textsc{Trec}\footnote{http://trec.nist.gov/data/web/12/gdeval.pl}.

\begin{figure}[!t]
	\centering
	\includegraphics[width=0.9\linewidth]{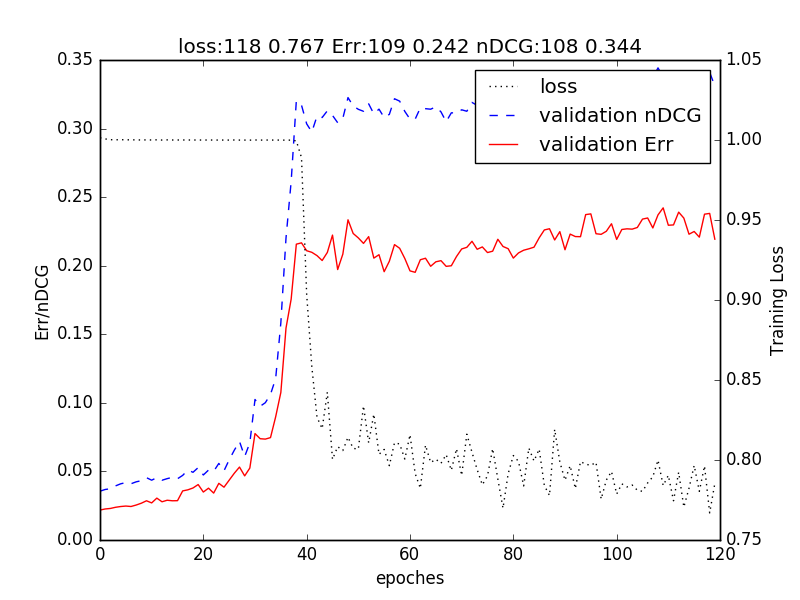} 
	\caption{The training loss, ERR@20 and nDCG@20 per iteration on validation data when training 
		on Web Track 2010--14. The x-axis denotes the iterations. The y-axis indicates the 
		ERR@20/nDCG@20 (left) and the loss (right).
		The best performance appears on 109th iteration with ERR@20=0.242. The lowest training loss (0.767) occurs after 118 iterations.}\label{fig.training}
\end{figure}

\paragraphHd{Training}
At each step, we perform Stochastic Gradient Descent (SGD) with a mini-batch of 32 triples.
For the purpose of choosing the triples,
we consider all documents that are judged with a label more relevant 
than \textit{Rel}\footnote{Judgments from \textsc{Trec} include junk pages
(\textit{Junk}), non-relevance (\textit{NRel}), relevance
(\textit{Rel}), high relevance (\textit{HRel}), key pages
(\textit{Key}) and navigational pages (\textit{Nav}).}
as \textit{highly relevant}, and put the remaining relevant documents into a
\textit{relevant} group.
To pick each triple, we sample a relevance group
with probability proportional to the number of documents in the group within the training set,
and then we randomly sample a document with the chosen label to serve as the positive document $d^+$. If the chosen group is
the highly relevant group, we randomly sample a document from the relevant group to serve as the negative document $d^-$.
If the chosen group is the relevant group, we randomly sample a non-relevant document
as $d^-$. This sampling procedure ensures that we differentiate between 
highly relevant documents
(i.e., those with a relevance label of \textit{HRel}, \textit{Key} or \textit{Nav}) 
and relevant documents (i.e., those are labeled as \textit{Rel}).
The training continues until a given number of iterations is reached.
The model is saved at every iteration. We use the model with the best ERR@20 on the validation set to make predictions.
Proceeding in a round-robin manner, we report test results on one year by
exploiting the respective remaining five years (250 queries) for training. From these 250 queries, 
we reserve 50 random queries as a held-out set for validation and hyper-parameter tuning, while the remaining 200 queries serve as the actual training set.

As mentioned, model parameters and training iterations are chosen by maximizing the ERR@20 on the validation set. The selected model is 
then used to make predictions on the test data.
An example of this training procedure is shown in Figure~\ref{fig.training}. 
There are four hyper-parameters that govern the behavior of the proposed \textit{PACRR-kwindow}
and \textit{PACRR-firstk}: the unified length of the document dimension $l_d$,
the k-max pooling size $n_s$, the maximum n-gram size $l_g$, and the number of 
filters used in convolutional layers $n_f$.
Due to limited computational resources, we determine 
the range of hyper-parameters to consider based on pilot experiments and 
domain insights. In particular, we evaluate $l_d\in [256,384,512,640,768]$,
$n_s\in [1,2,3,4]$, and $l_g\in [2,3,4]$. 
Due to the limited possible matching patterns given 
a small kernel size (e.g., $l_g=3$), $n_f$ is fixed to $32$.
For \textit{PACRR-firstk}, we intuitively desire to retain as much information
as possible from the input, and thus $l_d$ is always set to $768$.

\textit{DRMM} (\textit{DRMM}$_{LCH\times IDF}$), 
\textit{DUET}, \textit{MatchPyramid} and \textit{K-NRM} are trained under the same settings using
the hyperparameters described in their respective papers.
In particular,
as our focus is on the deep relevance matching model as mentioned in Section~\ref{sec.intro}, 
we only compare against DUET's local model, denoted as \textit{DUETL}.
In addition, \textit{K-NRM} is trained   
slightly different from the one described in~\cite{xiong2017end},
namely, with a frozen word embedding layer.
This is to guarantee its fair comparison with other models,
given that most of the compared models can be enhanced by co-training 
the embedding layers, whereas the focus here is the strength coming from 
the model architecture.
A fully connected middle layer with 30 neurons
is added to compensate for the reduction
of trainable parameters in \textit{K-NRM},
mirroring the size of DRMM's first fully connected layer.

All models are implemented with Keras~\cite{chollet2015keras}
using Tensorflow as backend, and are trained
on servers with multiple CPU cores. In particular, 
the training of \textit{PACRR}
takes 35 seconds per iteration on average,
and in total at most 150 iterations are trained
for each model variant.

\subsection{Results}\label{sec.result}

\begin{table*}[!t]
\centering
\ra
\resizebox{1\textwidth}{!}{%
 \begin{tabular}{@{}cc|c cc c|c cc cc cc c|c c@{}}
\toprule
Measure&Years&PACRR-firstk&Rank&PACRR-kwindow&Rank&DUETL&Rank&DRMM&Rank&MatchPyramid&Rank&K-NRM&Rank&QL&Rank\\
\midrule
\multirow{3}{*}{\shortstack{\textit{ERR@20}}}	
&wt12	&0.318~($mQ$)&2	&0.313~($MQ$)	&4	&0.281~($Q$)	&10	&0.289~($Q$)	&10	&0.227	&16	&0.258~($Q$)&12&0.177	&26\\
&wt13	&0.166~($DKQ$)	&3	&0.139~($Q$)	&14	&0.147~($Q$)	&12	&0.124	&25	&0.141~($q$)	&13	&0.134~($q$)&14&0.101	&38\\
&wt14	&0.221~($LMQ$)	&2	&0.208~($Q$)	&3	&0.179~($Q$)	&12	&0.193~($Q$)	&10	&0.176~($Q$)	&12	&0.201~($Q$)&8&0.131	&25\\
\cmidrule{2-16} 
\multirow{3}{*}{\shortstack{\textit{nDCG@20}}}
&wt12	&0.243~($DLMQ$)	&2	&0.250~($DLMQ$)	&2	&0.186~($Q$)	&11	&0.197~($Q$)	&8	&0.164~($Q$)	&16	&0.222~($Q$)&4&0.106	&39\\
&wt13	&0.295~($DLkQ$)	&3	&0.279~($DQ$)	&4	&0.248~($q$)	&11	&0.228	&20	&0.258~($Q$)	&7	&0.251~($Q$)&11&0.190	&36\\
&wt14	&0.339~($LMQ$)	&1	&0.331~($LMQ$)	&1	&0.267~($q$)	&11	&0.300~($Q$)	&6	&0.278~($Q$)	&10	&0.324~($Q$)&2&0.231	&23\\
\bottomrule

\end{tabular}
}
 \caption{ERR@20 and nDCG@20 on \textsc{Trec} Web Track 2012--14
 when re-ranking search results from \textit{QL}. 
The comparisons are conducted between two variants of PACRR and 
DRMM (D/d),
DUETL (L/l), MatchPyramid (M/m) and K-NRM (K/k). All methods are compared
against the \textit{QL} (Q/q) baseline. The upper/lower-case characters
in the brackets indicate a significant difference under
two-tailed paired Student's t-tests at 95\% or 90\% confidence levels 
relative to the corresponding approach.
 In addition, the relative ranks among all runs within the respective years
 according to ERR@20 and nDCG@20 are also reported directly after the absolute scores.
 }\label{tab.wrtql}
\end{table*}

\paragraphSc{RerankSimple}\label{sec.rerankql}
We first examine the proposed model by
re-ranking the search results from the \textit{QL} baseline on Web Track 2012--14.
The results are summarized in Table~\ref{tab.wrtql}. 
It can be seen that \textit{DRMM} 
can significantly improve \textit{QL} on WT12 and WT14, whereas \textit{MatchPyramid} fails on WT12 under ERR@20.
While \textit{DUETL} and \textit{K-NRM} can consistently outperform \textit{QL},
the two variants of \textit{PACRR} are the only models that can achieve significant improvements
at a 95\% significance level on all years under both ERR@20 and nDCG@20.
More remarkably, by solely re-ranking the search results from \textit{QL},
\textit{PACRR-firstk} can already rank within the top-3 participating systems on all three years
as measured by both ERR and nDCG. 
The re-ranked search results from \textit{PACRR-kwindow} also ranks within the top-5 
based on nDCG@20.
On average, both \textit{PACRR-kwindow} and \textit{PACRR-firstk}
achieve 60\% improvements over \textit{QL}.

\begin{table*}[!t]
	\centering
	\ra
	\resizebox{1\textwidth}{!}{%
		\begin{tabular}{@{}ccc|cccccc@{}}
			\toprule
			\multirow{1}{*}{}&\multirow{1}{*}{Measures}&Tested Methods&wt09&wt10&wt11&wt12&wt13&wt14\\
			\cmidrule{2-9}
			\multirow{10}{*}{\specialcell{average $\Delta$~measure score~over each year~(\%):\\ 
		  $\frac{\text{re-rank score}-\text{original score}}{\text{original score}}$}}&\multirow{5}{*}{ERR@20}
			&\textit{PACRR-firstk}&66\%~($DLK$)&362\%~($dm$)&43\%~($DLMK$)&76\%~($DLMK$)&37\%~($DLMK$)&41\%~($DLMK$)\\
			&&\textit{PACRR-kwindow}&70\%~($DLmK$)&393\%~($DlM$)&10\%~($LMK$)&83\%~($DLMK$)&21\%~($DLM$)&36\%~($DLMK$)\\
                        &&\textit{DUETL}& 80\%~($DMK$)& 316\%& 15\%~($DMK$)& 64\%~($M$)& 26\%~($DM$)& 19\%~($MK$)\\
			&&\textit{DRMM}&54\%~($LMK$)&315\%&11\%~($LMK$)&61\%~($M$)&5\%~($LMK$)&19\%~($MK$)\\
                        &&\textit{MatchPyramid}&65\%~($DL$)&313\%&2\%~($DLK$)&48\%~($DLK$)&29\%~($DLK$)&14\%~($DLK$)\\
                        &&\textit{K-NRM}&59\%~($DL$)&333\%&31\%~($DLM$)&63\%~($M$)&25\%~($DM$)&32\%~($DLM$)\\

			\cmidrule{2-9}
			&\multirow{5}{*}{nDCG@20}
			&\textit{PACRR-firstk}&69\%~($DLMK$)&304\%~($LM$)&56\%~($DLMK$)&100\%~($DLMK$)&31\%~($DLMK$)&31\%~($DLM$)\\
			&&\textit{PACRR-kwindow}&63\%~($DmK$)&345\%~($DLMK$)&27\%~($DLMK$)&113\%~($DLMK$)&23\%~($DLK$)&30\%~($DLM$)\\
			&&\textit{DUETL}& 62\%~($DMK$)& 237\%~($DK$)& 17\%~($DMK$)& 55\%~($DMK$)& 17\%~($DMK$)& 10\%~($DMK$)\\
			&&\textit{DRMM}&49\%~($LMK$)&274\%~($LMk$)&8\%~($LMK$)&70\%~($LMK$)&9\%~($LMK$)&15\%~($LK$)\\
                        &&\textit{MatchPyramid}&59\%~($DLk$)&232\%~($DK$)&1\%~($DLK$)&37\%~($DLK$)&21\%~($DLk$)&14\%~($LK$)\\
                        &&\textit{K-NRM}&52\%~($DLm$)&288\%~($dLM$)&36\%~($DLM$)&85\%~($DLM$)&19\%~($DLm$)&30\%~($DLM$)\\
			\midrule
			\multirow{10}{*}{\specialcell{\%~of runs that get better performance \\
					after re-ranking}}&\multirow{5}{*}{ERR@20}
			&\textit{PACRR-firstk}&94\%&95\%&\textbf{97\%}&92\%&\textbf{87\%}&\textbf{100\%}\\
			&&\textit{PACRR-kwindow}&\textbf{97\%}&\textbf{100\%}&47\%&\textbf{96\%}&65\%&76\%\\
			&&\textit{DUETL}& 94\% & 95\% & 61\% & 86\%  & 69\%& 59\%\\
			&&\textit{DRMM}&82\%&95\%&47\%&86\%&40\%&66\%\\
			&&\textit{MatchPyramid}& 85\% & 93\% & 40\% & 78\%  & 81\%& 59\%\\
			&&\textit{K-NRM}& 87\% & 95\% & 89\% & 82\%  & 67\%& 86\%\\
			\cmidrule{2-9}

			&\multirow{5}{*}{nDCG@20}
			
			&\textit{PACRR-firstk}&\textbf{94\%}&\textbf{100\%}&\textbf{100\%}&\textbf{100\%}&\textbf{92\%}&\textbf{93\%}\\
			&&\textit{PACRR-kwindow}&93\%&\textbf{100\%}&84\%&\textbf{100\%}&81\%&86\%\\
			&&\textit{DUETL}& 86\% & 93\% & 69\% & 92\% & 79\% & 59\%\\
			&&\textit{DRMM}&86\%&\textbf{100\%}&50\%&88\%&62\%&55\%\\
			&&\textit{MatchPyramid}& 76\% & 93\% & 39\% & 80\% & 81\% & 69\%\\
			&&\textit{K-NRM}&\textbf{94\%} & \textbf{100\%} & 97\% & 96\% & 81\% & \textbf{93\%}\\
			\bottomrule
		\end{tabular}
	}
			\caption{
		The average statistics when re-ranking all runs from the \textsc{Trec} Web Track 2009--14
		based on ERR@20 and nDCG@20.
		The average differences of the scores for individual runs are reported in the top portion.
		The comparisons are conducted between two variants of PACRR and 
		DRMM (D/d),
		DUETL (L/l), MatchPyramid (M/m)  and K-NRM (K/k). 
		The upper/lower-case characters
in parentheses indicate a significant difference under
two-tailed paired Student's t-tests at 95\% or 90\% confidence levels, respectively, 
relative to the corresponding approach.
The percentage of runs that show improvements in terms of a measure is summarized in the bottom
portion.}\label{tab.rerankstat}

\end{table*}

\paragraphSc{RerankALL}\label{sec.rerank}
In this part, we would like to further examine the performance of the proposed models
in re-ranking different sets of search results.
Thus, we extend our analysis to
re-rank search results from all submitted runs
from six years of the \textsc{Trec} Web Track ad-hoc task.
In particular,  we only consider the judged documents from 
\textsc{Trec}, which loosely correspond to top-20 documents in each run. 
The tested models make predictions for individual documents,
which are used to re-rank the documents within each submitted run.
Given that there are about 50 runs for each year, it is no longer
feasible to list the scores for each re-ranked run. 
Instead,
we summarize the results by comparing the performance of each run 
before and after re-ranking,
and provide statistics over each year to compare 
the methods under consideration in Table~\ref{tab.rerankstat}.
In the top portion of Table~\ref{tab.rerankstat},
we report the relative changes in metrics before and after re-ranking
in terms of percentages (``average $\Delta$ measure score'').
In the bottom portion, we report the percentage of systems whose results have increased after re-ranking.
Note that these results assess two different aspects:
the average $\Delta$ measure score in Table~\ref{tab.rerankstat}
captures the degree to which a model can improve an initial run, while
the percentages of runs indicate to what extent an improvement 
can be achieved over runs from different systems. In other words, the former measures the strength of the models, while the latter
measures the adaptability of the models.
Both \textit{PACRR} variants improve upon existing rankings by at least 10\% across different years.
Remarkably, in terms of nDCG@20, at least 80\% of the submitted runs are improved after re-ranking by the proposed models on individual years,
and on 2010--12, all submitted runs are consistently improved by \textit{PACRR-firstk}.
Moreover,
both variants of \textit{PACRR} can significantly outperform all baseline models
on at least three years out of the six years in terms of average improvement.
However,
it is clear that none of the tested models can make consistent improvements 
over all submitted runs across all six years.
In other words, there still exist
document pairs that are predicted contradicting to the judgments from \textsc{Trec}.
Thus, in the next part, we further investigate the performance in terms of prediction over 
document pairs.

\begin{table*}
\ra
	\centering
	\resizebox{1\textwidth}{!}{%
		\begin{tabular}{@{}ccc|cc|cccc@{}}
			\toprule
			\multirow{2}{*}{Label Pairs}& \multirow{2}{*}{Volume~(\%)}&\multirow{2}{*}{\# Queries}
			&\multicolumn{6}{c}{Tested Methods}\\
			\cline{4-9}
			&&&\textit{PACRR-firstk}&\textit{PACRR-kwindow}&\textit{DUETL}&\textit{DRMM}&\textit{MatchPyramid}&\textit{K-NRM}\\
			\midrule
			\textit{Nav}-\textit{HRel}&0.3\%&49&45.8\%&45.5\%& 45.2\%&48.2\%&47.3\%&51.6\%\\

			\textit{Nav}-\textit{Rel}&1.1\%&65&56.0\%~($m$)&56.3\%~($M$)& 54\%&57\%~($M$)&53.2\%~($D$)&57.4\%\\

			\textit{Nav}-\textit{NRel}&3.6\%&67&76.1\%~($DLMK$)&76.6\%~($DLMK$)& 67.1\%~($M$)&71.5\%~($M$)&64.7\%~($DLK$)&70.8\%~($M$)\\

			\textit{HRel}-\textit{Rel}&8.4\%&257&57.3\%&57.0\%& 55.5\%&55.8\%&52.8\%&56.1\%\\

			\textit{HRel}-\textit{NRel}&23.1\%&262&76.7\%~($DLMK$)&76.4\%~($DLMK$)& 68.4\%~($K$)&70.1\%~($MK$)&65.6\%~($DK$)&72.5\%~($DLM$)\\

			\textit{Rel}-\textit{NRel}&63.5\%&290&73.0\%~($DLMK$)&72.5\%~($DLMK$)& 63.9\%~($DMK$)&65.9\%~($LMK$)&61.4\%~($DLK$)&68.7\%~($DLM$)\\
			\midrule
			\multicolumn{3}{c|}{weighted average}&72.4\%&72.0\%&64.2\%&66.1\%&61.6\%&68.4\%\\
			\bottomrule
		\end{tabular}
	}
			\caption{Comparison among tested methods in terms of 
		accuracy when comparing document pairs with different labels. 
		The ``volume'' column indicates the percentage of occurrences
		of each label combination out of the total pairs.
		The ``\#~Queries'' column records the number of queries that include
		a particular label combination. 
		The comparisons are conducted between two variants of PACRR and 
		DRMM (D/d),
		DUETL (L/l), MatchPyramid (M/m) and K-NRM (K/k). 
		The upper/lower-case characters
in parentheses indicate a significant difference under
two-tailed paired Student's t-tests at 95\% or 90\% confidence levels, respectively, 
relative to the corresponding approach.
		In the last row, the average accuracy among different kinds of label combinations
		is computed, weighted by their corresponding volume.
	}\label{tab.labelpairstat}
\end{table*}

\paragraphSc{PairAccuracy}
The ranking of documents 
can be decomposed into rankings of document pairs as suggested 
in~\cite{Radinsky2011}. Specifically,
a model's retrieval quality can be examined by checking across a range of individual document pairs, namely,
how likely a model can assign a higher score for a more relevant document.
Thus, it is possible for us to compare different models over the same set of complete 
judgments, removing the issue of different initial runs.
Moreover, although ranking is our ultimate target,
a direct inspection of pairwise prediction results can indicate 
which kinds of document pairs a model succeeds at or fails on.
We first convert the graded judgments from \textsc{Trec} into ranked document pairs
by comparing their labels.
Document pairs are created among documents that 
have different labels. A prediction is counted as correct if it assigns
a higher score to the document from the pair that is labeled with a higher degree of relevance.
The judgments from
\textsc{Trec} contain at most six relevance levels, 
and we merge and unify the original levels from the six years into four grades,
namely,
\textit{Nav}, \textit{HRel}, \textit{Rel} and \textit{NRel}.
We compute the accuracy for each pair of labels.
The statistics are summarized in Table~\ref{tab.labelpairstat}.
The volume column lists the percentage of a given label combination out of 
all document pairs, and the \#~query column provides the number of queries for which the label combination exists.
In Table~\ref{tab.labelpairstat},
we observe that both \textit{PACRR} models
always perform better than all baselines
 on label combinations \textit{HRel} vs.\ \textit{NRel},
\textit{Rel} vs.\ \textit{NRel} and \textit{Nav} vs.\ \textit{NRel}, 
which in total cover 90\% of all document pairs.
Meanwhile, apart from \textit{Nav}-\textit{Rel},
there is no significant difference when distinguishing \textit{Nav} from other types.
\textit{K-NRM} and \textit{DRMM} perform better than the other two baseline models.


\subsection{Discussion} 
\label{sec:discussion}

\begin{figure*}[!t]
\centering
\includegraphics[trim=2cm 0cm 0cm 0cm, width=0.7\linewidth]{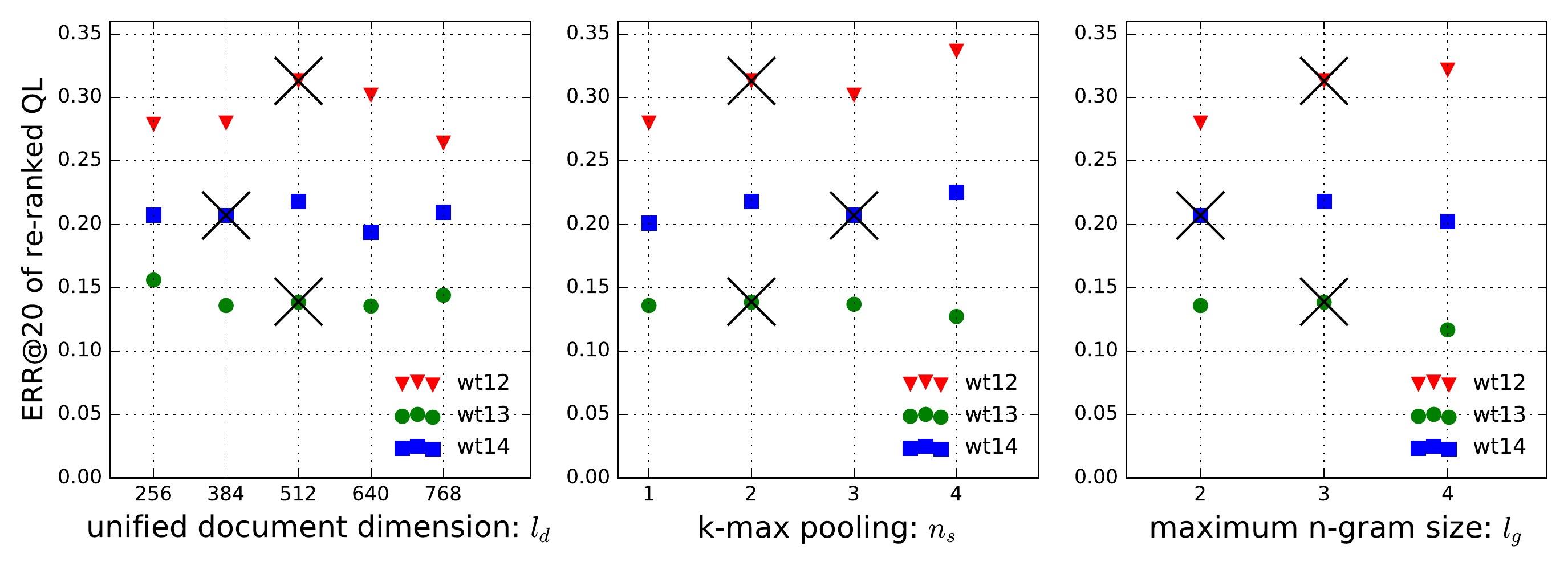} 
\caption{The ERR@20 of re-ranked \textit{QL} with \textit{PACRR-kwindow}
when applying different hyper-parameters: $l_d$, $n_s$ and $l_g$.
The x-axis reflects the settings for 
hyper-parameters, and the y-axis is the ERR@20. 
Crosses correspond to the selected models.}\label{fig.params}
\end{figure*}

\paragraphHdNospace{Hyper-parameters}
As mentioned, 
models are selected based on the ERR@20 over validation data.
Hence, it is sufficient to 
use a reasonable and representative validation dataset, rather than 
handpicking a specific set of parameter settings.
However, to gain a better understanding of the influence of different hyper-parameters,
we explore \textit{PACRR-kwindow}'s effectiveness when several hyper-parameters are varied.
The results when re-ranking \textit{QL} search results  
are given in Figure~\ref{fig.params}.
The results are reported based on the models with the
highest validation scores after fixing certain hyper-parameters. For example,
the ERR@20 in the leftmost figure is obtained when fixing $l_d$ to the values shown.
The crosses in Figure~\ref{fig.params} correspond to the models that were selected for use on the test data,
based on their validation set scores. It can be seen that the selected models are not necessarily
the best model on the test data, as evidenced by the differences between validation 
and test data results, but we consistently obtain scores within a reasonable margin.
Owing to space constraints, 
we omit the plots for \textit{PACRR-firstk}.

\paragraphHd{Choice between \textit{kwindow} and \textit{firstk} approaches}
As mentioned, both \textit{PACRR-kwindow} and 
\textit{PACRR-firstk}
serve to address the variable-length
challenge for documents and queries, and to make the training 
feasible and more efficient.
In general, if both training and test documents are known to be short enough to 
fit in memory, then \textit{PACRR-firstk} can be used directly.
Otherwise, \textit{PACRR-kwindow} is a reasonable choice to 
provide comparable results.
Alternatively, one can regard this choice as another hyper-parameter,
and make a selection based on held-out validation data.

\paragraphHd{Accuracy in \textsc{PairAccuracy}}
Beyond the observations in Section~\ref{sec.result},
we further examine the methods' accuracy over binary judgments by
merging the \textit{Nav}, \textit{HRel} and \textit{Rel} labels.
The accuracies become 
73.5\%, 74.1\% and 67.4\% for \textit{PACRR-kwindow}, \textit{PACRR-firstk},
and \textit{DRMM}, respectively.
Note that the manual judgments that
indicate a document as relevant or non-relevant relative to a given query 
contain disagreements~\cite{carterette2008here,voorhees2000variations}
and errors~\cite{alonso2012using}.
In particular,
a 64\% agreement (cf.\ Table 2~(b) therein) is observed over the inferred relative order among document pairs
based on graded judgments from six trained judges~\cite{carterette2008here}.
When reproducing \textsc{Trec} judgments,
Al-Maskari et al.~\cite{al2008relevance} reported a 74\% agreement (cf.\ Table 1 therein)
with the original judgments from \textsc{Trec}
when a group of users re-judged 56 queries on the \textsc{Trec}-8 document collections.
Meanwhile, Alonso and Mizzaro~\cite{alonso2012using} 
observed a 77\% agreement relative to judgments from \textsc{Trec}
when collecting judgments via crowdsourcing.
Therefore, the more than 73\% agreement achieved by both \textit{PACRR} methods
is close to the aforementioned agreement levels among different human assessors.
However, when distinguishing \textit{Nav}, \textit{HRel}, and \textit{Rel},
the tested models still fall significantly short of the human judges' agreement levels.
These distinctions are important
for a successful ranker, especially when measuring with graded metrics such as ERR@20 and nDCG@20.
Hence, further research is needed for better 
discrimination among relevant documents with different degrees of relevance.
In addition, as for the distinction between \textit{Nav} documents and 
\textit{Rel} or \textit{HRel} documents,
we argue that since \textit{Nav} actually indicates that a document mainly satisfies
a navigational intent, this makes
such documents qualitatively different from \textit{Rel} and \textit{HRel} documents. 
Specifically, a \textit{Nav} is more relevant for a user with 
navigational intent, whereas for other users it may in some cases be less useful than 
a document that directly includes highly pertinent information content.
Therefore, we hypothesize that further improvements
can be obtained by introducing a classifier for user intents,
e.g., navigational pages, 
before employing neural IR models.


\section{Conclusion}
\label{sec:conclusion}
In this work, we have demonstrated the importance of preserving
positional information for neural IR 
models by incorporating domain insights into the proposed \textit{PACRR} model.
In particular, \textit{PACRR} captures term dependencies and proximity
through multiple convolutional layers with different sizes.
Thereafter, following two max-pooling layers,
it combines salient signals over different query terms
with a recurrent layer.
Extensive experiments show that \textit{PACRR}
substantially outperforms four state-of-the-art neural IR models on \textsc{Trec} Web Track ad-hoc datasets and can dramatically improve search results when used as a re-ranking model.


\newpage
\bibliography{kai} 

\begin{thebibliography}{23}
\expandafter\ifx\csname natexlab\endcsname\relax\def\natexlab#1{#1}\fi

\bibitem[{Al-Maskari et~al.(2008)Al-Maskari, Sanderson, and
  Clough}]{al2008relevance}
Azzah Al-Maskari, Mark Sanderson, and Paul Clough. 2008.
\newblock Relevance judgments between trec and non-trec assessors.
\newblock In \emph{Proceedings of the 31st annual international ACM SIGIR
  conference on Research and development in information retrieval}, pages
  683--684. ACM.

\bibitem[{Alonso and Mizzaro(2012)}]{alonso2012using}
Omar Alonso and Stefano Mizzaro. 2012.
\newblock Using crowdsourcing for trec relevance assessment.
\newblock \emph{Information Processing \& Management}, 48(6):1053--1066.

\bibitem[{Carterette et~al.(2008)Carterette, Bennett, Chickering, and
  Dumais}]{carterette2008here}
Ben Carterette, Paul~N Bennett, David~Maxwell Chickering, and Susan~T Dumais.
  2008.
\newblock Here or there: {Preference Judgments for Relevance}.
\newblock In \emph{Advances in Information Retrieval}, pages 16--27. Springer.

\bibitem[{Chapelle et~al.(2009)Chapelle, Metlzer, Zhang, and
  Grinspan}]{Chapelle2009ERR}
Olivier Chapelle, Donald Metlzer, Ya~Zhang, and Pierre Grinspan. 2009.
\newblock \href {https://doi.org/10.1145/1645953.1646033} {Expected reciprocal
  rank for graded relevance}.
\newblock In \emph{Proceedings of the 18th ACM conference on Information and
  knowledge management}, CIKM '09, pages 621--630, New York, NY, USA. ACM.

\bibitem[{Chollet et~al.(2015)}]{chollet2015keras}
Fran\c{c}ois Chollet et~al. 2015.
\newblock Keras.
\newblock \url{https://github.com/fchollet/keras}.

\bibitem[{Guo et~al.(2016)Guo, Fan, Ai, and Croft}]{guo2016deep}
Jiafeng Guo, Yixing Fan, Qingyao Ai, and W~Bruce Croft. 2016.
\newblock A deep relevance matching model for ad-hoc retrieval.
\newblock In \emph{Proceedings of the 25th ACM International on Conference on
  Information and Knowledge Management}, pages 55--64. ACM.

\bibitem[{Hochreiter and Schmidhuber(1997)}]{Hochreiter1997}
Sepp Hochreiter and J\"{u}rgen Schmidhuber. 1997.
\newblock \href {https://doi.org/10.1162/neco.1997.9.8.1735} {Long short-term
  memory}.
\newblock \emph{Neural Computation}, 9(8):1735--1780.

\bibitem[{Huang et~al.(2013)Huang, He, Gao, Deng, Acero, and
  Heck}]{Huang:2013:LDS:2505515.2505665}
Po-Sen Huang, Xiaodong He, Jianfeng Gao, Li~Deng, Alex Acero, and Larry Heck.
  2013.
\newblock \href {https://doi.org/10.1145/2505515.2505665} {Learning deep
  structured semantic models for web search using clickthrough data}.
\newblock In \emph{Proceedings of the 22Nd ACM International Conference on
  Information \& Knowledge Management}, CIKM '13, pages 2333--2338, New York,
  NY, USA. ACM.

\bibitem[{Hui et~al.(2017)Hui, Yates, Berberich, and de~Melo}]{hui2017position}
Kai Hui, Andrew Yates, Klaus Berberich, and Gerard de~Melo. 2017.
\newblock Position-aware representations for relevance matching in neural
  information retrieval.
\newblock In \emph{Proceedings of the 26th International Conference on World
  Wide Web Companion}, pages 799--800. International World Wide Web Conferences
  Steering Committee.

\bibitem[{Huston and Croft(2014)}]{DBLP:conf/cikm/HustonC14}
Samuel Huston and W.~Bruce Croft. 2014.
\newblock \href {https://doi.org/10.1145/2661829.2661894} {A comparison of
  retrieval models using term dependencies}.
\newblock In \emph{Proceedings of the 23rd {ACM} International Conference on
  Conference on Information and Knowledge Management, {CIKM} 2014, Shanghai,
  China, November 3-7, 2014}, pages 111--120. {ACM}.

\bibitem[{J{\"a}rvelin and
  Kek{\"a}l{\"a}inen(2002)}]{jarvelin2002cumulatedNDCG}
Kalervo J{\"a}rvelin and Jaana Kek{\"a}l{\"a}inen. 2002.
\newblock Cumulated gain-based evaluation of ir techniques.
\newblock \emph{ACM Transactions on Information Systems (TOIS)},
  20(4):422--446.

\bibitem[{Kalchbrenner et~al.(2014)Kalchbrenner, Grefenstette, and
  Blunsom}]{kalchbrenner2014convolutional}
Nal Kalchbrenner, Edward Grefenstette, and Phil Blunsom. 2014.
\newblock A convolutional neural network for modelling sentences.
\newblock \emph{arXiv preprint arXiv:1404.2188}.

\bibitem[{Metzler and Croft(2005)}]{metzler2005markov}
Donald Metzler and W~Bruce Croft. 2005.
\newblock A markov random field model for term dependencies.
\newblock In \emph{Proceedings of the 28th annual international ACM SIGIR
  conference on Research and development in information retrieval}, pages
  472--479. ACM.

\bibitem[{Mikolov et~al.(2013)Mikolov, Sutskever, Chen, Corrado, and
  Dean}]{mikolov2013distributed}
Tomas Mikolov, Ilya Sutskever, Kai Chen, Greg~S Corrado, and Jeff Dean. 2013.
\newblock Distributed representations of words and phrases and their
  compositionality.
\newblock In \emph{Advances in neural information processing systems}, pages
  3111--3119.

\bibitem[{Mitra et~al.(2017)Mitra, Diaz, and Craswell}]{mitra2017learning}
Bhaskar Mitra, Fernando Diaz, and Nick Craswell. 2017.
\newblock Learning to match using local and distributed representations of text
  for web search.
\newblock In \emph{Proceedings of WWW 2017}. ACM.

\bibitem[{Mitra et~al.(2016)Mitra, Nalisnick, Craswell, and
  Caruana}]{mitra2016dual}
Bhaskar Mitra, Eric Nalisnick, Nick Craswell, and Rich Caruana. 2016.
\newblock A dual embedding space model for document ranking.
\newblock \emph{arXiv preprint arXiv:1602.01137}.

\bibitem[{Ounis et~al.(2006)Ounis, Amati, Plachouras, He, Macdonald, and
  Lioma}]{ounis06terrier}
Iadh Ounis, Gianni Amati, Vassilis Plachouras, Ben He, Craig Macdonald, and
  Christina Lioma. 2006.
\newblock Terrier: A high performance and scalable information retrieval
  platform.
\newblock In \emph{Proceedings of the OSIR Workshop}, pages 18--25.

\bibitem[{Pang et~al.(2016)Pang, Lan, Guo, Xu, and
  Cheng}]{DBLP:journals/corr/PangLGXC16}
Liang Pang, Yanyan Lan, Jiafeng Guo, Jun Xu, and Xueqi Cheng. 2016.
\newblock \href {http://arxiv.org/abs/1606.04648} {A study of matchpyramid
  models on ad-hoc retrieval}.
\newblock \emph{CoRR}, abs/1606.04648.

\bibitem[{Radinsky and Ailon(2011)}]{Radinsky2011}
Kira Radinsky and Nir Ailon. 2011.
\newblock \href {https://doi.org/10.1145/1935826.1935850} {Ranking from pairs
  and triplets: Information quality, evaluation methods and query complexity}.
\newblock In \emph{Proceedings of the Fourth ACM International Conference on
  Web Search and Data Mining}, WSDM '11, pages 105--114, New York, NY, USA.
  ACM.

\bibitem[{Tao and Zhai(2007)}]{tao2007exploration}
Tao Tao and ChengXiang Zhai. 2007.
\newblock An exploration of proximity measures in information retrieval.
\newblock In \emph{Proceedings of the 30th annual international ACM SIGIR
  conference on Research and development in information retrieval}, pages
  295--302. ACM.

\bibitem[{Voorhees(2000)}]{voorhees2000variations}
Ellen~M Voorhees. 2000.
\newblock Variations in relevance judgments and the measurement of retrieval
  effectiveness.
\newblock \emph{Information processing \& management}, 36(5):697--716.

\bibitem[{Wang et~al.(2017)Wang, Yu, Zhang, Gong, Xu, Wang, Zhang, and
  Zhang}]{wang2017irgan}
Jun Wang, Lantao Yu, Weinan Zhang, Yu~Gong, Yinghui Xu, Benyou Wang, Peng
  Zhang, and Dell Zhang. 2017.
\newblock Irgan: A minimax game for unifying generative and discriminative
  information retrieval models.
\newblock \emph{arXiv preprint arXiv:1705.10513}.

\bibitem[{Xiong et~al.(2017)Xiong, Dai, Callan, Liu, and Power}]{xiong2017end}
Chenyan Xiong, Zhuyun Dai, Jamie Callan, Zhiyuan Liu, and Russell Power. 2017.
\newblock End-to-end neural ad-hoc ranking with kernel pooling.
\newblock \emph{arXiv preprint arXiv:1706.06613}.

\end{thebibliography}
\bibliographystyle{emnlp_natbib}

\end{document}